\documentclass[showpacs,amsmath,amssymb,prb,aps,twocolumn]{revtex4}
\usepackage{graphicx}
\begin{document}
\title {``Phase Diagram" of the Spin Hall Effect}
\author {E. M. Hankiewicz $^{1,2}$}
\email{hankiewicze@missouri.edu}
\author {G. Vignale $^1$}
\affiliation{$^1$ Department of Physics and Astronomy, University
of Missouri, Columbia, Missouri 65211, USA\\ $^2$ Department of
Physics, Fordham University, Bronx, New York 10458, USA}
\date{\today}
\begin{abstract}
We obtain analytic formulas for the frequency-dependent spin-Hall
conductivity of a two-dimensional electron gas (2DEG) in the
presence of impurities, linear spin-orbit Rashba interaction, and
external magnetic field perpendicular to the 2DEG. We show how
different mechanisms (skew-scattering, side-jump, and spin
precession)  can be brought in or out of focus by changing
controllable parameters such as frequency, magnetic field, and
temperature. We find, in particular, that the d.c. spin Hall
conductivity vanishes in the absence of a magnetic field, while a
magnetic field restores the skew-scattering and side-jump
contributions proportionally to the ratio of magnetic and Rashba
fields.
\end{abstract}
 \maketitle
The spin-Hall effect (SHE), i.e. the generation
of a transverse spin current in response to a d.c. electric field
\cite{Perel,Hirsch99,Zhang00,Sinova04,Murakami03,Raimondi05} has
attracted much attention recently, particularly after a series of
experiments~\cite{Kato04,Sih05,Wunderlich05,Stern06} culminating
in the observation of the SHE at room temperature~\cite{Stern06}.
On the theoretical front, however, there remains considerable
uncertainty as to the physical origin of the SHE, which appears to
depend on an intricate interplay of three processes:  (i) the
skew-scattering (SS) due to spin-orbit interaction (SOI) between electrons and impurities,
(ii) the side-jump (SJ) (due to the non-canonical character of the
physical position and velocity variables in the presence of SOI
with impurities), and  (iii) the spin precession caused by spin
non-conserving terms in the band structure, among which we include
the linear Rashba SOI generated by an external electric field. To
these we may add the influence of an external magnetic field,
which tends to lock the spins in a fixed direction, thus reducing
the importance of spin precession. A first principles theory that
include all of these effects on equal footing is very complicated.
To our knowledge, the diagrammatic approach by Tse and Das
Sarma~\cite{Sarm06} comes closest to fulfilling the order, and yet
it does not include magnetic field or frequency.  However, these
diagrammatic calculations are very difficult to follow in detail
and do not lead to an intuitive understanding of the striking
nonadditive behavior of impurity and band structure effects.

Our goal in this paper is to present the ``phase diagram" of the
SHE, i.e.  to clarify in which range of experimentally
controllable parameters one should expect the dominance of each
mechanism mentioned above, and how the crossovers between
different regimes occur. We do this for a 2DEG with Rashba SOI and
a magnetic field perpendicular to the plane. The two parameters
that are most easily controlled in an experiment are (i) the
frequency $\omega$ of the electric field and (ii) the magnitude of
the perpendicular magnetic field, whose strength is characterized
by the resonance frequency $\omega_0$. Accordingly, we plot our
``phase-diagram" in the $\omega \tau-\omega_0/\alpha_R k_F$ plane,
where $\tau$ is the electron impurity scattering time and
$\alpha_R k_F$ is the magnitude of the effective magnetic field
due to Rashba spin-orbit coupling for electrons at the Fermi
wavevector $k_F$. Throughout the paper we assume
$\omega,1/\tau,\alpha_R k_F,\omega_0 \ll E_F$, i.e. all energy
scales are smaller than the Fermi energy.

\begin{figure}[thb]
\vskip 0.0 in
\includegraphics[width=3.2in]{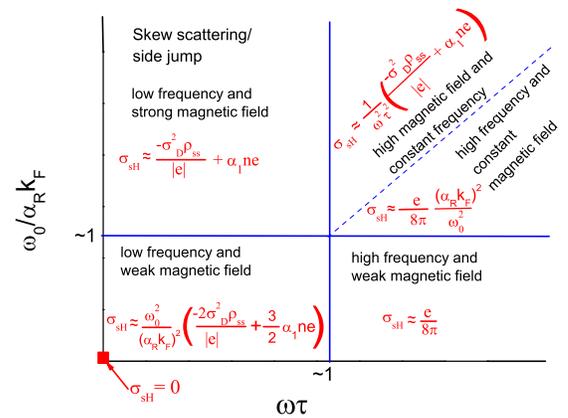}
\caption{Different regimes of spin-Hall effect in the $\omega\tau
- \omega_0/\alpha_R k_F$ plane. $\sigma_D$ is Drude conductivity,
$\rho_{ss}$ is skew scattering. Spin-orbit interactions and
mobility are fixed. See discussion in the text.}
\label{PhaseDiagram}
\end{figure}

 Our qualitative conclusions are shown in Fig.~(\ref{PhaseDiagram}).
In contrast to previous calculations \cite{Sarm06}, we found that
the d.c. limit ($\omega\tau\to0$) of the spin Hall conductivity is zero in the
presence of spin precession and in the absence of a magnetic field ($\omega_0=0$).  As the magnetic field
increases, both SS and SJ contributions increase with the ratio of
the magnetic to the Rashba field ($\omega_0/\alpha k_F$), thus restoring the values
they would have had at zero frequency and zero magnetic field in the absence of spin precession.
 In high-mobility samples the SS mechanism is
the dominant mechanism \cite{Hankiewicz05,HankiewiczPRL06}
overcoming the SJ contribution, which has an opposite sign for
attractive impurity potential.  However,  the SJ mechanism  could
well dominate in low-mobility samples.  As discussed in
Ref.~\cite{HankiewiczPRL06}, the mobility can be  controlled to
some extent by changing the temperature $T$, and this allows one
in principle to tune between SS and SJ  contributions. For this
reason we use the label SS/SJ for the left side of diagram, where
$\omega \tau \ll 1$. In the a.c. regime ($\omega\tau\gg1$) and for
low magnetic field impurities become irrelevant, leaving room for
the intrinsic SHE with ``universal" conductivity $e/8\pi$.
\cite{Sinova04}
 Finally, in the a.c. regime and at high magnetic field  the spin Hall conductivity declines
 to zero in  different manners depending on whether  $\omega \tau $ or the magnetic field
is  kept constant as shown in Fig.~(\ref{PhaseDiagram}).

In what follows we describe the main points of our new theoretical approach,
which enables us to calculate the spin Hall conductivity in different regions of the parameter space
and to derive simple analytic formulas describing the crossovers between different regimes.

Our model is defined by the hamiltonian
\begin{eqnarray}\label{Hamiltonian}
\hat H &=&\sum_{i=1}^N \left\{\frac{\hat{p}^2_{i}}{2 m^*} + V(\hat
{\vec r}_i)+\frac{\alpha_R }{\hbar}\left(\hat p_{iy}\hat S_{ix}-
\hat p_{ix}\hat S_{iy}\right)\right.\nonumber\\
&+&\left.\frac{\alpha_1}{\hbar}\left(\hat p_{ix}\nabla_yV(\hat
{\vec r}_i) -\nabla_xV(\hat{\vec r}_i)\hat p_{iy}\right)\hat
S_{iz} +  \omega_0 \hat S_{iz}\right\}~
\end{eqnarray}
Besides the kinetic energy and the usual electron-impurity
potential $V(r)$, we have {\it two} distinct spin-orbit couplings.
 The $\alpha_1$ coupling between the electrons and the impurities conserves the $z$-component of the spin and is responsible for SS and
  the SJ effects.    The $\alpha_R$ coupling --  also known as Rashba coupling --
 creates a momentum-dependent magnetic field in the plane,
 which breaks the conservation of $S_z$ and causes spin precession.
 This term is responsible, under appropriate conditions, for the intrinsic contribution to SHE.  Finally, we have
  included a magnetic field perpendicular to the plane.
The Zeeman splitting, $\omega_0$,  and the frequency $\omega$ of
the a.c. electric field (not shown in $\hat H$) are the two
control knobs in terms of which our ``phase diagram" will be
plotted.

The skew-scattering effect is easily described in the Boltzmann
equation formalism\cite{Hankiewicz05}, but it is difficult to treat in the diagrammatic approach\cite{Sarm06}.
On the other hand, spin precession effects are easily included in the diagrammatic formalism,
but are problematic in the Boltzmann equation formalism (the distribution function becomes a $2 \times 2$ matrix).
 We get the best of two worlds by combining the two approaches in the following manner.
   First we notice that the skew-scattering collision term in the Boltzmann equation is formally equivalent to
   the imposition  of a ``spin-electric field" $E^z_y$, which accelerates up-spin and down-spin electrons
   (``up"  and ``down"  are defined with respect to a $z$-axis) in opposite directions along the $y$ axis,
   perpendicular to the flow of the charge current ($x$).   The problem is now ``reduced" to calculating
   the $z$-spin current $j^z_y$ which flows along the $y$ axis in response to the spin electric field
   $E^z_y$ in the same direction.  This can be done with the help of the standard diagrammatic formalism, including both electron-impurity scattering and spin precession, but {\it not} the skew-scattering processes; for the skew scattering has already ``done its job" by producing  the spin electric field  $E^z_y$ (in other words, we work to first order in $\alpha_1$).
 Thus we have
\begin{equation}\label{jss}
j^z_y \vert_{ss}= \sigma^z_{yy}E^z_y
\end{equation}
where $\sigma^z_{yy}$ is the longitudinal spin conductivity calculated in the absence
of skew-scattering (or side-jump)
effects.  On the other hand, $E^z_y$ has the well-known expression
\begin{equation}\label{EZY}
E^z_y = \rho_{ss} j_x
\end{equation}
where  $j_x$ is the current density in the $x$ direction and
$\rho_{ss}\equiv -m/ne^2\tau_{ss}$  is the skew-scattering
resistivity,  calculated from the Boltzmann collision
integral.~\cite{Hankiewicz05}  An explicit expression for the
skew-scattering rate $1/\tau_{ss}$ is given in Eq.(29) of
Ref.~\cite{Hankiewicz05}:   notice that it is proportional to
$\alpha_1$ and positive. \footnote{spin-precession corrections to
Eq.~(\ref{EZY}) are of order of $\alpha_R^2\alpha_1$ and we omit
them as small.} $j_x$ can be expressed as $\sigma_D E_x$, where
$E_x$ is the electric field in the $x$ direction and $\sigma_D$ is
the Drude conductivity of the electron gas. Therefore
Eq.~\ref{jss} becomes:
\begin{equation}
j^z_y\vert_{ss} = \sigma^z_{yy}\rho_{ss}\sigma_D E_x,
\end{equation}
from which we extract the first important result of this paper,
namely the expression for the skew-scattering contribution to the spin Hall conductivity $\sigma^{SH}$,
\begin{equation}\label{sigmash}
\sigma^{SH}_{ss} = \sigma^z_{yy}\rho_{ss}\sigma_D
\end{equation}
The  dependence of this formula on spin-precession rate, frequency
of the a.c. field, and magnetic field will be obtained below.  In
particular we will show that $\sigma^{SH}_{ss}$ vanishes at low
magnetic field (spin-precession regime), and it recovers the zero-precession value
($\sigma_D^2\rho_{ss}\hbar/e$) at high magnetic field.

We will then consider separately the remaining contributions
$\sigma^{SH}_{sj}$ and $\sigma^{SH}_{R}$ to the spin Hall
conductivity. They are most efficiently analyzed in terms of the
Kubo formula for the spin-Hall conductivity. The SJ contribution,
similar to SS term vanishes at low magnetic field (spin-precession
regime), and recovers its value in high magnetic fields.
The remaining intrinsic contribution  is easily calculated by
standard diagrammatics (including vertex corrections), and leads
to the well known $e/8\pi$ result in the  appropriate a.c. regime.

{\it Skew-scattering} -- As discussed above, the central role in
calculating $\sigma^{SH}_{ss}$ is played by the longitudinal
spin-channel conductivity $\sigma^z_{yy}$.  The formal expression for
the real part of $\sigma^z_{yy}$ is
\begin{equation}\label{longitudinal_spin_conductivity}
\sigma^z_{yy} (\omega)= - \frac{4 n  e}{m^{*2}\hbar} \frac{ \Im m
\langle\langle \hat S_z \hat p_y; \hat S_z \hat
p_y\rangle\rangle}{\omega}
\end{equation}
where $\hat p_y$, $\hat S_z$ are momentum and spin operators for a single electron
\footnote{Substitution of operator for total system $\sum_{i=1}^N
\hat p_{iy}S_{iz}$ by sum of one-electron operators $Np_yS_z$ is
justified because we omit electron-electron interactions.}, the
double bracket is the usual notation for the spin-current-spin
current response function, $n$ is the areal density of the electron gas.
Our objective is to calculate this
Kubo formula including elastic electron-impurity scattering in the
Born approximation, spin precession due to a Rashba spin-orbit
interaction and a magnetic field
 along the $z$-axis, but neglecting spin-orbit interactions with the impurities.
The Feynman diagrams for $\sigma^z_{yy} (\omega)$, to the desired
level of accuracy, are shown in Fig.~2.
\begin{figure}[t]
\vskip 0.27 in
\includegraphics[width=3.2in]{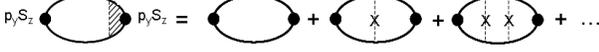}
\caption{Diagrammatic representation of ladder approximation to
calculate the spin-channel conductivity.}
\end{figure}
 The solid lines represent $2\times 2$ Green's functions  including Rashba spin-orbit coupling,
 magnetic field, and  a diagonal disorder self-energy $(i/2 \tau) {\rm sgn}(\omega)$, where  $1/\tau$
  is the elastic scattering rate:
\begin{eqnarray}\label{Green_function}
\hat G(p,\omega) &=&\frac{\omega-\epsilon_p+\vec h_p\hat {\vec S}
+\frac{i}{2\tau}{\rm sgn} (\omega)}{\left(\omega-\epsilon_p
+\frac{i}{2\tau}{\rm sgn} (\omega)\right)^2-h_p^2/4}.
\end{eqnarray}
Here $\vec h_p =(\alpha_R k_y,-\alpha_R k_x,\omega_0)$ combines
the Rashba in-plane field ($x$- and $y$-components) and the
magnetic field along the $z$ axis in a single effective field. The
dashed lines are electron-impurity interactions (averaged over
disorder), and there are spin-current operators $p_yS_z$ at the
vertices. The empty bubble, calculated by the standard procedure
with $\alpha_R k_F$, $\omega_0$, and $1/\tau$ all much smaller
than the Fermi energy is
\begin{eqnarray}\label{bubble}
\sigma^z_{yy} (\omega)\vert_{\rm bubble} \simeq
\frac{\hbar}{e}\frac{\sigma_D}{1+(\omega\tau)^2} \frac {1+
\omega_0^2 \tau^2 }{1+ \Omega^2 \tau^2}
\end{eqnarray}
where $\sigma_D = ne^2\tau/m^*$ is the Drude conductivity and
$\Omega =\sqrt{(\alpha_R k_F)^2+\omega_0^2}$ is the spin
precession frequency. Eq.~(\ref{bubble}) suggests that the spin
channel conductivity is similar to the ordinary Drude
conductivity, only somewhat reduced by precession about the
in-plane Rashba field. The inclusion of vertex corrections changes
the result drastically. The sum of the ladder diagrams produces
 a ``dressed" spin-current vertex  $\hat \Lambda^z_y$ of  the form:
\begin{eqnarray}\label{vertex_funcRashb}
\hat \Lambda^z_y = \hat p_y \hat S_z &-& \frac{\alpha_R k_F^2
[(\omega_0\tau)\hat S_{x}-(1-i\omega\tau)\hat S_{y}]
}{\alpha_R^2k_F^2\tau-2\Omega^2\tau(1-i\omega\tau)}.
\end{eqnarray}
Including $\hat \Lambda^z_y$ in
the calculation of the spin-channel conductivity we get
 \begin{eqnarray}\label{SpinChannelConductivity.omega}
 \sigma^z_{yy}(\omega) =
 \frac{2\hbar \sigma_D}{e}\frac{\cos^2\delta(1+\cos^2\delta)}{(1+\cos^2\delta)^2
 + 4\omega^2\tau^2}
\end{eqnarray}
where $\cos \delta = \frac{\omega_0}{\Omega}$ is the cosine of the
angle between the external magnetic field and the full effective
magnetic field $\vec \Omega$. While
Eq.~(\ref{SpinChannelConductivity.omega})  reduces to the
empty-bubble result in the limit $\omega_0 \gg \alpha_R k_F$, its
most striking feature is that it vanishes identically (i.e. at all
frequencies) in the opposite limit $\omega_0 \ll \alpha_R k_F$. In
other words, the spin channel conductivity is zero in the absence
of an external magnetic field, as long as an infinitesimal spin
precession rate $\alpha_R k_F$ is present! Accordingly, the skew
scattering spin-Hall conductivity from  Eq.~(\ref{sigmash}) is
\footnote{Actually, this is not the complete formula for the
skew-scattering contribution (which is quite complicated),
 but a simplified version of it, which  is valid in the limit of $\alpha_Rk_F \tau \gg 1$.
 However, even in the opposite regime $\alpha_Rk_F \tau \ll 1$ this formula remains qualitatively correct,
 in the sense that it exhibits the correct scaling behaviors in the limits of weak and high magnetic field,
 low and high frequency.  Notice that contributions of order $\omega^2/\Omega^2$, which we have omitted,
 give  finite SS at finite frequency and zero magnetic field, but are negligible
  in comparison to the larger intrinsic contribution.}
 \begin{eqnarray}\label{Skewscattering.omega}
\sigma^{SH}_{ss} =\frac{2\hbar \sigma^2_D\rho_{ss}}{e}
\frac{\cos^2\delta(1+ \cos^2\delta)}{(1+ \cos^2\delta)^2 +
4\omega^2\tau^2}~
\end{eqnarray}
which vanishes in the absence of an external magnetic field and
recovers the value $\sigma^2_D\rho_{ss}\hbar/e$ for strong
magnetic fields. The vanishing of the spin-conductivity is a
peculiar feature of the linear Rashba model, in which the spin
current $\hat p_y\hat S_z$ is proportional to the time derivative
of $\hat S_y$: $\hat p_y \hat S_z=-\dot{\hat S}_y/\alpha_R$.  The
expectation value of a time derivative must vanish at zero
frequency.
 This is the same reason which
causes the vanishing of the Rashba spin-Hall conductivity in the
d.c. limit.

{\it Side-jump} -- In a recent paper~\cite{HankiewiczPRL06} we
have shown how to identify the SJ contribution in the Kubo linear
response formalism. We perturb the hamiltonian
~(\ref{Hamiltonian}) with a uniform electric field of frequency
$\omega$ in the $x$ direction. Then after a series of
manipulations, which have been described in Ref.~
\cite{HankiewiczPRL06}, we arrive at the following Kubo formula for
the spin Hall conductivity:
\begin{eqnarray}\label{linear.response.1}
&&\sigma^{SH}_{yx}(\omega)=\frac{n e}{ i
m^*\omega}\left(\langle\langle \hat p_{y}\hat S_{z};\frac{\hat
{p}_x}{m^*}-\frac{\alpha_R}{\hbar}\hat S_y\rangle\rangle\right)
\\ &&+  \frac{\alpha_1 n e}{2i
m^*\omega}\left(\frac{4\langle\langle \hat p_{y}\hat
S_{z};\nabla_y V(\hat {\vec
r})\hat{S}_{z}\rangle\rangle}{\hbar^2}-\langle\langle \nabla_x
V(\hat {\vec r});\hat{p}_x\rangle\rangle \right)~\nonumber
\end{eqnarray}
The first term on the right hand side produces SS as well as
possible ``intrinsic" contributions which are discussed in next
section. The second line of this equation, which explicitly shows
a dependence on $\alpha_1$, is responsible for the SJ contribution
\footnote{We omitted corrections of order $\alpha_R^2\alpha_1$.},
denoted $\sigma^{SH}_{sj}$. We now focus on this contribution.
Using the Heisenberg equation of motion for the momentum operator
$\dot{\hat p}_{ix} =-\nabla_xV(\hat {\vec r}_i)$ in zero order in
$\alpha_1$ (because SJ terms are already explicitly linear in
$\alpha_1$), we rewrite the SJ contribution as
\begin{eqnarray}\label{sj.1}
\sigma^{SH}_{sj}(\omega) = \frac{e\alpha_1 n}{2i
m^*\omega}\left[\langle\langle
\dot{\hat{p}}_{x};\hat{p}_x\rangle\rangle  - \frac{4\langle\langle
\hat p_{y}\hat S_{z}; \dot{\hat
p}_{y}\hat{S}_{z}\rangle\rangle}{\hbar^2} \right]
\end{eqnarray}
In the first term in the square brackets we apply the standard
rule of integration, which allows us to replace
$\dot{\hat{p}}_{x}/i\omega$ by $-\hat p_x$ and rewrite it (its
real part) as:
\begin{eqnarray}\label{sj.1a}
\frac{\alpha_1 e n}{2im^*\omega}\langle\langle
\dot{\hat{p}}_{x};\hat{p}_x\rangle\rangle= \frac{-\alpha_1 n
e}{2m^*}\langle\langle \hat{p}_x,\hat{p}_x\rangle\rangle
=\frac{\alpha_1ne}{2(1+\omega^2\tau^2)},~
\end{eqnarray}
where the last equality follows from the well-known form of the current-current
response function in a weakly disordered system.   The second
term in the square brackets  of Eq.~(\ref{sj.1}) can be rewritten as follows:
\begin{eqnarray}\label{sj.2}
-\frac{1}{i\omega}\langle\langle\hat p_{y}\hat S_{z};\dot{\hat
p}_{y}\hat{S}_{z}\rangle\rangle = -\frac{1}{i\omega}\langle\langle
\hat p_{y}\hat S_{z}; \frac{d}{dt}\left({\hat
p}_{y}\hat{S}_{z}\right) - \hat {p}_{y}\dot{\hat S}_{z}
\rangle\rangle ~~
\end{eqnarray}
Applying again the integration formula, the first term of
Eq.~\ref{sj.2} yields:
\begin{eqnarray}\label{sj.2b}
\frac{-2en\alpha_1}{m^*} \langle\langle\hat p_{y}\hat S_{z}; \hat
p_{y}\hat{S}_{z}\rangle\rangle = \frac{e \alpha_1
n}{2}\frac{\cos^2\delta(1+\cos^2\delta)^2}{(1+\cos^2\delta)^2+4\omega^2\tau^2}\nonumber\\
\end{eqnarray}
In the absence of spin precession (i.e., for $\omega_0 \gg
\alpha_Rk_F$, $cos\delta \simeq 1$) and for zero frequency the
contributions~(\ref{sj.1a}), (\ref{sj.2b})  add up to the ``usual"
 SJ conductivity  $\alpha_1ne$ of Ref.\cite{HankiewiczPRL06}. In the opposite limit of  $\omega_0 \ll \alpha_Rk_F$ ($cos\delta =0$) this term vanishes for the
same reason as $\sigma^{SH}_{ss}$.   The second term on the rhs of
Eq.~(\ref{sj.2}) can be rewritten as $\langle\langle \hat S_y;\hat
{p}_{y}(\hat{\vec p} \cdot \hat{\vec S}) \rangle\rangle$  (we use
the equations of motion $\dot {\hat S}_z = \alpha_R \hat {\vec p}
\cdot \hat {\vec S}$  and $\hat p_y \hat S_z=-\dot{\hat
S}_y/\alpha_R$).
\begin{figure}[t]
\includegraphics[width=3.2in]{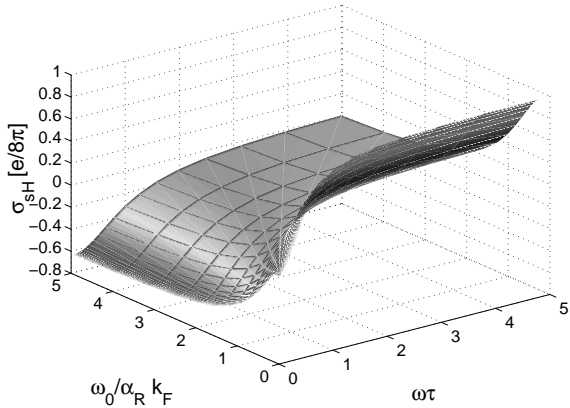}
\caption{Behavior of spin-Hall conductivity as a function of
$\omega\tau$ and $\omega_0/\alpha_R k_F$ and carrier concentration
n$_{2D}= 2\times 10 ^{12}$cm$^{-2}$, $\mu =$1m$^2$/Vs, $m
=0.067m_e$, $\alpha_1=$0.053nm$^2$,$\alpha_R k_F=$10meV. For
calculation of skew-scattering, we assumed square potential
characterized by $\tau/\tau_{ss}=0.002$ like in
Ref.~\cite{Hankiewicz05}}
\end{figure}
In the limit of zero frequency and zero magnetic field this
reduces to $\langle\langle \hat p_y \hat S_y;\hat p_y \hat
S_y\rangle \rangle= (\hbar^2/4) \langle\langle \hat p_x;\hat
p_x\rangle \rangle$, which exactly cancels the contribution
from~(\ref{sj.1a}) and causes the SJ contribution to vanish. For
general magnetic fields and frequencies the total
$\sigma^{SH}_{sj}$  is finally given by the formula\footnote{Again, we present
 only a simplified version of the complete formula, subject to restrictions explained in footnote $^{18}$.}
\begin{eqnarray}\label{sj.2d}
\sigma^{SH}_{sj}(\omega) = \frac{\alpha_1
ne}{2}\left[\frac{\cos^2\delta}{1+\omega^2\tau^2}+\frac{2\cos^2\delta(1+\cos^2\delta)}
{(1+\cos^2\delta)^2+4\omega^2\tau^2}\right],~
\end{eqnarray}
which in the d.c. limit simplifies to
\begin{eqnarray}\label{sj.2e}
\sigma^{SH}_{sj}(\omega)=
\frac{\alpha_1ne\cos^2\delta}{2}\left(\frac{3+\cos^2\delta}{1+\cos^2\delta}\right).
\end{eqnarray}
Summarizing, the side-jump conductivity in  the d.c. limit
($\omega \tau \ll 1$) grows from zero at $\omega_0=0$ to the
``full" value ($\alpha_1ne$) at high magnetic fields, while it
tends to zero in high frequency limit.

 {\it Rashba contribution} --
Evaluating the first line of Eq.~(\ref{linear.response.1})  in the
absence of skew-scattering and side-jump, which we have already
taken into account, leads to the well-known result
\cite{Inoue04,Halperin04} $\sigma^{SH}_{R}(\omega) = \frac{e}{8
\pi}\frac{(\omega \tau)^2}{(\omega\tau)^2+1/4}$, which vanishes at
$\omega=0$ and tends to the ``ballistic" limit $e/8 \pi$ for
$\omega \gg 1/\tau$. We have found that the behavior of $
\sigma_{R}^{SH}(\omega)$ as a function of external perpendicular
magnetic field and frequency is given by:
\begin{eqnarray}\label{finalhz}
\sigma^{SH}_R(\omega) =
  \frac{e}{8\pi}\frac{\alpha_R^2k_F^2}{\Omega^2}\frac{4\omega^2\tau^2}{(1+\cos^2\delta)^2+4\omega^2\tau^2},
\end{eqnarray}
which is a decreasing function of magnetic field.

{\it Conclusions} -- The results of our analysis are summarized in
Fig.~3 which shows the full spin Hall conductivity $\sigma_{SH}$,
including Rashba, skew-scattering and side-jump terms, as a
function of two variables, frequency and magnetic field, for
realistic values of the parameters.  The figure shows the smooth
crossovers between different regimes and should be useful to
experimentalists attempting to extricate the various components of
this still quite intriguing effect.

{\it Acknowledgements.} We thank H. A. Engel for useful
discussions. This work was supported by NSF Grant No. DMR-0313681.


\end{document}